\documentclass[prl,twocolumn,showpacs,preprintnumbers,amsmath,amssymb]{revtex4}
\usepackage{graphicx}
\usepackage{dcolumn}
\usepackage{bm}
\usepackage[dvipsnames]{xcolor}
\begin{document}
\title{Competing magnetic correlations across the ferromagnetic quantum critical point in the Kondo system CeTi$_{1-x}$V$_x$Ge$_3$: $^{51}$V NMR as a local probe}
\author{M. Majumder$^{1}$\footnote{Present address: Experimental Physics VI, Center for Electronic Correlations and Magnetism, University of Augsburg, Germany}}
\email{mayukh.cu@gmail.com}
\author{W. Kittler$^2$}
\author{V. Fritsch$^{2,3}$}
\author{H. v. L{\"o}hneysen$^{2,4}$}
\email{Hilbert.loehneysen@kit.edu}
\author{H. Yasuoka$^1$}
\author{M. Baenitz$^1$}
\email{Michael.Baenitz@cpfs.mpg.de}

\affiliation{$^1$Max Plank Institute for Chemical Physics of Solids, 01187 Dresden, Germany}
\affiliation{$^2$Physics Institute, Karlsruhe Institute of Technology, 76131-Karlsruhe, Germany}
\affiliation{$^3$Experimental Physics VI, Center for Electronic Correlations and Magnetism, Institute of Physics, University of Augsburg, 86135 Augsburg, Germany}
\affiliation{$^4$Institute for Solid State Physics, Karlsruhe Institute of Technology, 76012 Karlsruhe, Germany}

\date{\today}

\begin{abstract}

$^{51}$V nuclear magnetic resonance (NMR) and magnetization studies on CeTi$_{1-x}$V$_x$Ge$_3$ have been performed to explore the evolution from the ferromagnetic ($x = 0.113$) to the antiferromagnetic Kondo lattice state ($x = 1$), with focus on the emergence of a possible ferromagnetic quantum critical point (FMQCP) at $x_c \approx 0.4$. From the temperature dependence of the nuclear spin-lattice relaxation rate, $1/T_1T$, and the Knight shift, \textit{K}, for $x=0.113$ and $x=1$ a considerable competition between ferro- and antiferromagnetic correlations is found. Around the critical concentration ($x = 0.35, 0.405$) quantum-critical spin fluctuations entail weak antiferromagnetic spin fluctuations admixed with ferromagnetic spin fluctuations. The FMQCP in CeTi$_{1-x}$V$_x$Ge$_3$ therefore is not purely ferromagnetic in nature.

\end{abstract}

\maketitle

\section{introduction}

The generic Doniach diagram of heavy-fermion materials is based on the competition of Ruderman-Kittel-Kasuya-Yosida (RKKY) interaction between localized magnetic moments, and Kondo interaction between moments and conduction electrons. It has been derived for antiferromagnetic (AF) and isotropic RKKY-type of exchange interaction~\cite{Doniach}. For 4$f$ systems this competition is a prime scenario for the emergence of a quantum phase transition with a critical point (QCP) at zero temperature. The tuning of AF correlated 4$f$ systems by an external control parameter (e.g., chemical composition, hydrostatic or uniaxial pressure, magnetic field) towards an AFQCP has been demonstrated for many materials\cite{review,Gegenwart08,Si10}. The more recent search for ferromagnetic (FM) quantum critical points (FMQCP) among 4$f$ and 3$d$ systems~\cite{Brando} has provided a number of examples. However, due to the interplay between FM and AF correlations and, even more importantly, the low-lying quasiparticle excitations in metallic systems, these systems often undergo transitions to other phases and thus appear to "avoid" a bona-fide FMQCP. For correlated Yb systems FM order is rare and so far YbNi$_4$P$_2$ is the only FM 4$f$ system (with a Curie temperature $T_C$ of 170 mK) which could be tuned towards the FMQCP~\cite{Steppke13}. Among the Ce systems, CeRuPO~\cite{Krellner07}, CeRu$_2$Ge$_2$~\cite{Suellow99}, CeAgSb$_2$~\cite{Sidorovat03}, and CeNiSb$_2$~\cite{Sidorov05} exhibit FM order but application of hydrostatic pressure induces long-range AF order thus avoiding an FMQCP. In CeFePO~\cite{Lausberg12} and CePd$_{1-x}$Rh$_x$~\cite{Westerkamp09} a glass-like "Kondo cluster" state forms which likewise impedes tuning towards an FMQCP. 

NMR measurements were carried out on a number of those systems to probe locally the magnetic fluctuations and unveil their nature (AF vs. FM fluctuations) by temperature and field scaling of the spin-lattice relaxation rate, $1/T_1T$, and the NMR shift, $K$. In particular, NMR is able to qualitatively disentangle finite-$q$ (AF) and $q=0$ (FM) excitations~\cite{Baenitz13,Sarkar13}. At the same time, NMR provides information about the degree of disorder in alloy systems which might affect the nature of the quantum critical point. For example, in the heavy-fermion compound CeFePO where long-range order is absent, $^{31}$P NMR gave clear evidence for FM correlations admixed to AF correlations~\cite{Bruning08} whereas in the structural homologue CeRuPO the existence of stable long range FM order was demonstrated by $^{31}$P NMR measurements~\cite{Krellner07}. 

Recently, Kittler \textit{et al.} succeeded to tune the Curie temperature of the FM Kondo-lattice system CeTiGe$_3$ ($T_C \approx$ 14 K) monotonically down to $T$ = 0 indicating a possible FMQCP upon V substitution for Ti~\cite{Kittler13,Thesis}. $T_C$ decreases linearly with increasing V concentration, the structure type (hexagonal perovskite $P6_3/ mmc$, see Fig. 6) is preserved and no crossover to other phases occurs. Upon V substitution the $c/a$ ratio is reduced~\cite{Kittler13}. The single-ion anisotropy of the Ce$^{3+}$ ions arising from the effect of the crystal electric field (CEF) on the 4$f$ moments, changes drastically with increasing V content: while the easy direction is parallel to the $c$ axis for CeTiGe$_3$, it is perpendicular to $c$ for CeVGe$_3$~ \cite{Thesis,Kittler}. The end member on the V-rich side, CeVGe$_3$, is an antiferromagnetic Kondo system with a N\'{e}el temperature $T_N$ of about $6$ K and a Kondo temperature of $T_K \approx 10$ K~\cite{Kittler}. It should be noted that in contrast to most systems discussed above the V substitution of Ti in CeTiGe$_3$ is not isoelectronic but differs by one $d$ electron. Recent studies on CeTiGe$_3$ suggest an avoided FMQCP and a splitting of the $T_C(p)$ line into a wing-like structure~\cite{Udhara18} as observed in other clean ferromagnets~\cite{Brando}, with the complication of several intermediate AF phases occurring in this system. 

\section{Experimental details}

CeTi$_{1-x}$V$_x$Ge$_3$ alloys crystallize in the hexagonal perovskite (BaNiO$_3$-type) structure (space group $P6_3/mmc$) without any structural phase transitions across the entire V concentration range. The samples ($x$ = 0.113, 0.35, 0.405 and 1) used in this experiment were cut from single crystals~\cite{Thesis} and ground to powder. The $^{51}$V NMR spectra were taken by the transient (pulsed) NMR technique with a commercial NMR spectrometer (TecMag Apollo). The NMR line profiles were obtained by integration of the spin-echo amplitude in the time domain with sweeping an external field at constant frequency. By simulating the experimental line profile, the isotropic and anisotropic (from the V local symmetry an axial distribution of shift is expected, see Fig. 6) Knight shifts ($K_{iso}$ and $K_{ax}$) have been extracted. The spin-lattice relaxation rate $1/T_1$ was measured by the saturation recovery method where the recovery of nuclear magnetization was fitted to a stretched exponential function (see Fig. 11).

In general, NMR probes the hyperfine field at the nuclear site originating from the electron magnetic moment which has static and dynamic components. The $^{51}$V nucleus (natural abundance 100 \%) has a nuclear spin of $I = 7/2$, hence a quadrupolar splitting is expected under an electric-field gradient (EFG). However, the absence of such satellites indicates a small EFG at the V nuclear site. We have also checked by simulation that even if there is a quadrupolar effect, that is less than 100 kHz, does not contribute to the temperature dependence of the shift parameters. This is originated by the fact that $^{51}$V nucleus has a small quadrupole moment (0.051 barns) and that the EFG usually is screened by conduction electrons in a metal, so the $^{51}$V $\nu_Q$ effect has been ignored in our study. The temperature dependent static part of the hyperfine field can be measured by the NMR (Knight) shift, $K(T)$, which is related to the electron spin susceptibility, $\chi(T)$, by $K(T) = (A_{hf}/N\mu_B)\chi(T) + K_0$, where $A_{hf}$ is the hyperfine coupling constant between the nuclear and the electron spins, and $K_0$ is a temperature independent contribution. In metals, $K_0$ arises from both the orbital interaction associated with non-$s$ electrons and the Fermi contact interaction from $s$ conduction electrons ($K_0 = K_{orb} + K_{ce}$). For non-cubic materials like CeTi$_{1-x}$V$_x$Ge$_3$, the Knight shift is anisotropic reflecting the anisotropy of the spin susceptibility. In this case, one observes a powder pattern of the NMR spectrum arising from the different orientation of the crystallites with respect to applied field direction. For the present uniaxial case, we can extract values of $K_\parallel$ ($B_0$ is parallel to the symmetry axis) and $K_\perp$ ($B_0$ perpendicular to the symmetry axis) from the singularities of the spectrum. Then isotropic and axial Knight shifts, $K_{iso}$ and $K_{ax}$, respectively, are given by
\begin{equation}
K_{iso} = (K_\parallel +2 K_\perp)/3, K_{ax} = (K_\parallel-K_\perp)/3.
\end{equation}
Note that in the present case $K_\parallel$ corresponds to $K$ parallel to the $c$ axis and $K_\perp$ to $K$ perpendicular to the $c$ axis.

It is to be noted that we have found a considerable change in the magnitude of $A_{hf}$ with doping (listed in the Table I). Such a variation can be due to the change of lattice constant, along with the change of electronic structure (due to electron doing in the present case) as well. Replacing Ti$^{4+}$ by V$^{5+}$ introduces carriers (doping) into the system while the structure type is preserved. It reduces the $c/a$ ratio and changes the single-ion anisotropy of the Ce$^{3+}$ ion. The competition between these two effects might result in the observed considerable change in $A_{hf}$.

The fluctuating part of the hyperfine field is related to the spin-lattice relaxation rate $1/T_1$ given by,
\begin {equation}
1/T_1 \propto T \sum_q  |A_{hf}(q)|^2\chi^{\prime\prime}_\perp(q, \omega_n)/\omega_n.
\end {equation}	 		   
where the sum extends over the wave vectors $q$ within the first Brillouin zone, $\chi^{\prime\prime}_\perp(q, \omega_n$) is the imaginary part of the transverse dynamical electron spin susceptibility and $\omega_n$ is the Larmor frequency for NMR. $A_{hf}(q)$ is the $q$-dependent hyperfine coupling constant.

The recovery of the nuclear magnetization, $M(t)$, in these inhomogeneoussly broadened cases generally follows a stretched exponential form, then the recovery data have been fitted to $M(t)=(1-exp(-t/T_1)^\beta)$, where t is the time after a saturation of nuclear magnetization and $\beta$ is the stretch exponent related to the distribution of the $T_1$ value across the spectrum.

\section{results and discussions}
Before discussing the compounds close to the FMQCP ($x = 0.35$ and 0.405) we present the results for the parent compounds (AF, $x = 1.0$ and FM, $x = 0.113$).

\begin{figure}[h]
{\centering {\includegraphics[width=8.5cm]{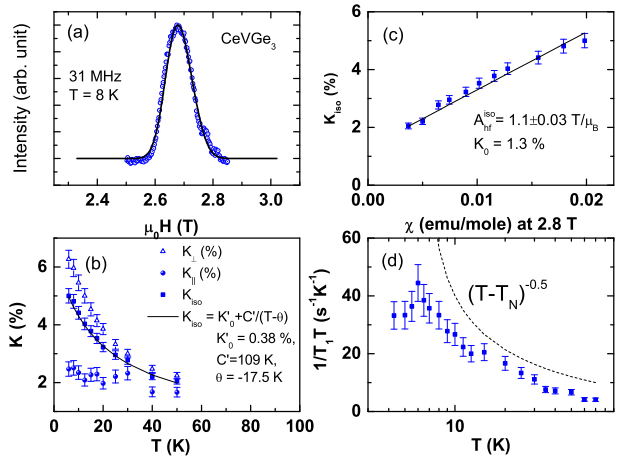}}\par} \caption{NMR data for $x$=1 sample. (a) $^{51}$V NMR spectra measured at 31 MHz at 8 K, solid line is a fit of a Gaussian line profile to the data; (b) temperature dependence of $K_{\perp}$ (open triangles), $K_\parallel$ (closed circles) and the resultant $K_{iso}$, solid line is a fit of a Curie-Weiss law to the $K_{iso}$ data; (c) $K_{iso}$ versus $\chi$ measured at 2.8 T, solid line is a linear fit of the data; (d) temperature dependence of $1/T_1T$, dashed line is an expected curve of the SCR theory for itinerant AF magnet.} \label{structure}
\end{figure}

CeVGe$_3$ ($x = 1$) is a heavy-fermion itinerant antiferromagnet with $T_K \approx$ 10 K and $T_N \approx$ 6 K. The $^{51}$V NMR spectra have been taken at a fixed frequency of 31.0 MHz. The field-swept $^{51}$V NMR spectra at 8 K, the temperature dependence of $K_{iso}$ and the relaxation rate divided by temperature, $1/T_1T$, are depicted in Fig. 1(a), (b) and (d), respectively. In order to obtain the isotropic hyperfine coupling constant $A_{hf}^{iso}$, the observed $K_{iso}$ data are plotted as a function of the magnetic susceptibility ($\chi$) with the temperature as an implicit parameter in Fig. 1(c). From the slope and the intercept $A_{hf}^{iso}$ and $K_{orb}$ are obtained as 1.1 T/Ce-$\mu_B$ and 1.3\%, respectively. Here, $A_{hf}^{iso}$ indicates that if a Ce 4$f$ moment is polarized by one $\mu_B$, the V nucleus is subjected to an isotropic hyperfine field of 1.1 T along the field direction.

$K_{iso}$, hence the static susceptibility $\chi$, has a strong temperature dependence, implying that the wave-vector-dependent susceptibility $\chi(q)$ has not only a peak at wave vector $q = Q$ but also must have a large FM $q = 0$ component. As to the $1/T_1T$ vs $T$ plot in Fig.1(d), the self-consistent renormalization (SCR) theory for a weak itinerant AF predicts a $1/T_1T \propto 1/\sqrt{T-T_N}$ divergence at $T_N$~\cite{Moriya85}. Although there is an anomaly at $T_N$ in $1/T_1T$, the experimental data cannot be described by the above relation. This is also an evidence for the existence of FM fluctuations which is suppressed by an external field around $T_N$.

For the other end member of the series we have taken the $x = 0.113$ sample which undergoes FM ordering at $T_C \approx$ 9 K in zero field (as determined from the sharp maximum in the specific heat~\cite{Thesis}). The $^{51}$V NMR spectrum shown in Fig.2(a) exhibits a large anisotropy ($K_{\perp}/K_\parallel \simeq 0.26$ at 20 K) which is consistent with the anisotropy in the magnetic susceptibility reported for single crystals~\cite{Thesis,Inamdar14}. As will be described below, the anisotropy tends to vanish towards the FM QCP but increases again in CeVGe$_3$ where the anisotropy is reversed to yield $K_{\perp}/K_\parallel \simeq 2.37$ at 8 K, again in accord with the susceptibility measurements~\cite{Kittler13,Thesis}.

\begin{figure}[h]
{\centering {\includegraphics[width=8.5cm]{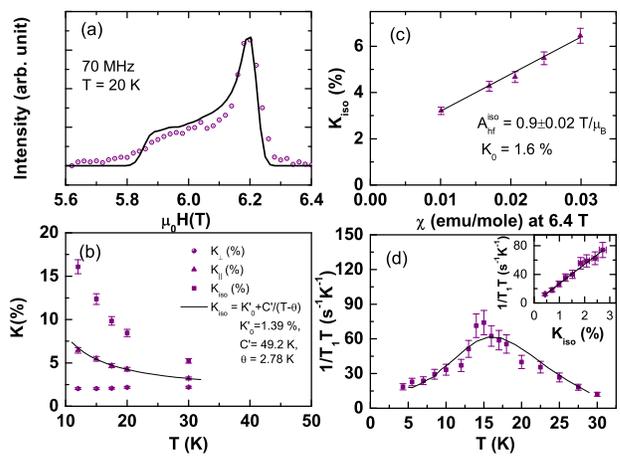}}\par} \caption{NMR data for the $x$=0.113 sample. (a) $^{51}$V NMR spectra measured at 70 MHz at 20 K, (b) temperature dependence of $K_{\perp}$, $K_\parallel$, and the calculated $K_{iso}$, (c) $K_{iso}$(\%) versus $\chi$ measured at 2.8 T, (d) temperature dependence of $1/T_1T$. The solid line is a fit of Eq. (3) to the data (for details see main text), the inset shows $1/T_1T$ vs $K_{iso}$.} \label{structure}
\end{figure}

The temperature dependence of $K_{iso}$ extracted from $K_{\parallel}$ and $K_{\perp}$ is shown in Fig. 2(b), and that of $1/T_1T$ in Fig. 2(d). Although $1/T_1T$ scales with $K_{iso}^2$ (Korringa law) in simple metals, $1/T_1T$ for $x=0.113$ depends linearly on $K_{iso}$ as shown in the insert of Fig. 2(d). This behavior can be described by the SCR theory for a weak itinerant ferromagnet in the frame of SCR theory. In the SCR theory, $1/T_1T$ under an external field is given by\cite{Moriya56,Moriya85,Moriya63}
\begin {equation}
 (1/T_1T) = \kappa\chi/(1+\chi^3B_0^2P),
\end {equation}
which reproduces very well the experimental $1/T_1T$(T) shown in Fig. 2(d) for $B_0$ = 6.25 T. Here, $P$ is a constant related to the area of the Fermi surface of the itinerant electrons, and $\kappa$ is related to the spin-fluctuation parameter, $T_0$. In the paramagnetic state
\begin {equation}
1/T_1T \simeq 3\hbar\gamma_n^2A_{hf}K_{iso}/16\pi\mu_BT_0.
\end {equation}
The slope of $1/T_1T$ vs. $K_{iso}$ in the inset of Fig. 2(d) yields the value of $T_0$ = 21 K which is rather close to $T_C$. The deviation from 1 of the ratio $T_C/T_0$ is a measure of the degree of itinerancy of the 4$f$ electrons~\cite{Moriya56,Moriya85,Moriya63,Majumder10}. $T_C \approx T_0$ indicates that the enhancement of $\chi(q)$ is not confined to $q = 0$ but rather extends to larger $q$ values. This situation is reminiscent of but opposite to that of $x=1$ described above, where $\chi(q$) extends from the AF wave vector $Q$ towards $q$ = 0. Therefore, we conclude that there is a considerable mixture of FM and AF correlations in both end materials.

\begin{figure}[h]
{\centering {\includegraphics[width=8.5cm]{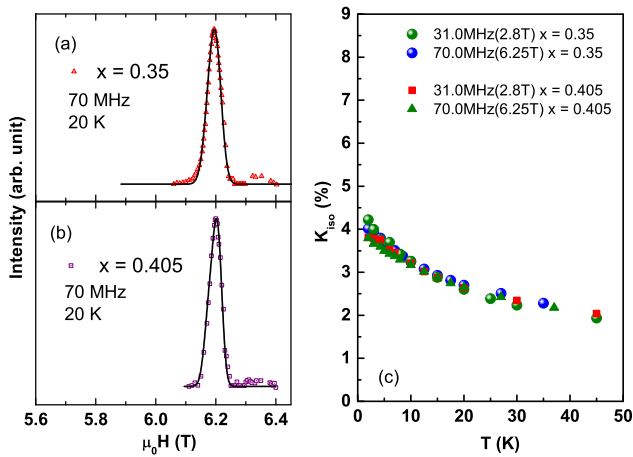}}\par} \caption{ $^{51}$V NMR spectra measured at 70 MHz and 20 K for (a) $x$ = 0.35 and (b) 0.405, (c) $K_{iso}$ for $x$ = 0.35 and 0.405 sample as a function of temperature in two different fields.} \label{structure}
\end{figure}


We now discuss $^{51}$V NMR data for the samples with $x = 0.35$ and 0.405 which are close to the proposed FMQCP. Here the NMR measurements were performed under different applied magnetic fields to elucidate the properties in low fields. The spectra shown in Fig. 3 (a, b) for $x = 0.35$ and $x = 0.405$, are less anisotropic compared to those for $x = 1.0$ and 0.113. The temperature dependence of $K_{iso}$ at 2.76 T and 6.25 T is shown in Fig. 3(c). The change of characteristics in the spin-fluctuations can easily be seen from Fig. 12, where $1/T_1T$ have been plotted as a fucntion of tempetaure for all the different doping concertrations. The temperature dependences of the relaxation rate $1/T_1$ in different fields are shown in Fig.4(a) and (b) for $x=0.35$ and $x=0.405$, respectively. $1/T_1$ shows a crossover from a low-temperature Korringa-like process ($1/T_1 \propto T$) to a temperature-independent process above a temperature, $T^*$ which depends on field. The SCR theory has been successfully adopted to understand thermodynamic properties of weakly or nearly itinerant FM and AF systems. This model has been extended to cover cases where the spin fluctuations have large amplitudes and their significant $q$ components extend over the entire $q$ space of the first Brillouin zone. This extension of the SCR theory, "temperature-induced-local-moment" (TILM) model, predicts that the amplitude of local spin fluctuations, $<S_L^2>$, increases rapidly at low temperatures and saturates at a certain temperature $T^*$~\cite{Moriya78}. Above $T^*$, the thermodynamic properties are governed by the transverse component (local-moment type) of the fluctuations and the susceptibility shows Curie-Weiss behavior. As described above, the $1/T_1$ data of CeTi$_{1-x}$V$_x$Ge$_3$ around $x \approx 0.4$ are in accord with TILM model. It has been shown that exchange-coupled local-moment fluctuations bring about a temperature independent $1/T_1$ a magnitude given by the correlation time $\tau_c$ associated with the local spin fluctuations ~\cite{Yasuoka}:
\begin {equation}
(1/T_1)_{TILM} = (A_{hf}/\hbar)^2 \dfrac{(2\pi)^{1/2}<S_L^2>}{3}\cdot\tau_c
\end {equation}
where $(1/T_1)_{TILM}$ is the temperature independent value observed above $T^*$, and $<S_L^2>$ is the amplitude of the local spin density. Using experimental values of $(1/T_1)_{TILM}$ and effective moments (obtained from the Curie-Weiss fit) assumed to be the same as $<S_L^2>$, $\tau_c$ is evaluated and shown in Fig. 4(c) as a function of magnetic field. $\tau_c$ for both samples around the FMQCP ($x$ = 0.35 and 0.405) increases linearly with field, indicating that the characteristic frequency of the spin fluctuations decreases with external field, i.e., the samples approach the field-polarized state. 

An possible microscopic reason for a constant $1/T_1$ value may be the suppression of Kondo interaction with increasing temperature, with a crossover from T-linear Fermi-liquid-like relaxation rate. Because $T_K$ which is around 10 K in CeTi$_{1-x}$V$_x$Ge$_3$, the high-temperature relaxation process may be due to fluctuations of purely local moments associated with the thermal quenching of the Kondo interaction. The relaxation rate of this type of process can easily be calculated using the hyperfine coupling constant and an exchange frequency estimated from the Weiss constant\cite{Moriya56}. The calculated $1/T_1$ is around $2\times10^3$ sec$^{-1}$, which is about one order of magnitude larger than the experimental values. Also, if we assume observed $T^*$ is associated with $T_K$ one would expect that the $T^*$ should be decreased with increasing filed that is opposite to the experimental findings. Therefore, we believe TILM is more plausible scenario to explain the experimental results for the samples close to the quantum critical point.

For a detailed discussion, we plot $T_1T$ which is a measure of the inverse $\chi^{\prime\prime}(q, \omega_n)$ (cf. Eq. 2) as a function of temperature in Fig. 5(a) and (b) for $x=0.35$ and $x=0.405$. For both samples $1/T_1T$ above $T^*$ follows the CW law with the Weiss constant $\theta$ increasing with applied magnetic field (figure 5(c)). The enhancement of $\theta$ simply illustrates the tendency to a saturated paramagnet with increasing field. The fact that $\theta$ for the lowest field for both samples is very close to zero signals quantum critical spin-fluctuations. Note that the field dependent Curie constants correspond  to the inverse of $\tau_c$ (cf. Eq. 5).    

\begin{figure}[h]
{\centering {\includegraphics[width=8.5cm]{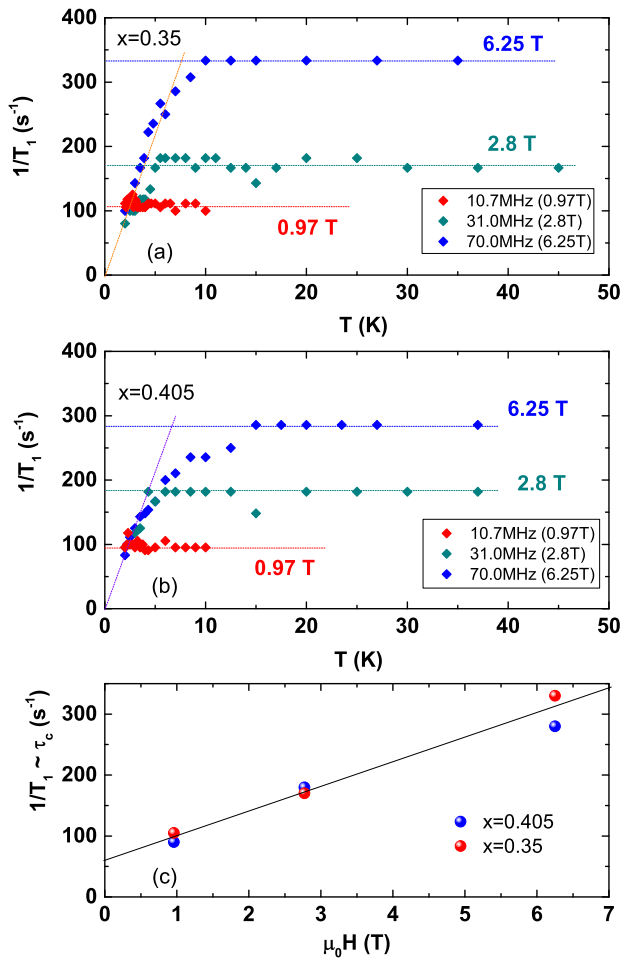}}\par} \caption{(a) and (b): $1/T_1$ vs. temperature T for $x$ = 0.35 and 0.405 at different fields, (c): $\tau_c$ (correlation time of the spin fluctuations) as a function of applied field. The dotted and solid lines are for guide to the eye.} \label{structure}
\end{figure}


As shown in Fig. 4(a) and (b), at temperatures well below $T^*$ the nuclear relaxation is governed by the Korringa process where interacting electron-hole excitations are the primary source of the magnetic excitations. Treating the interaction of the quasiparticles in the frame of the random phase approximation (RPA) the modified Korringa relation can be expressed as $S_0/T_1TK_{spin}^2 = K(\alpha)$, with $S_0 = (\hbar/4\pi k_B)(\gamma_e/\gamma_n)^2$. $K(\alpha) = <(1 - \alpha_0)^2/(1 - \alpha_q)^2)>_{FS}$ with  $\alpha_q = \alpha_0 [\chi_0(0,q)/\chi_0(0,0)]$ where $\chi_0(\omega,q)$ is the magnetic susceptibility, and $< ...>_{FS}$ indicates the $q$ average over the Fermi surface. $K(\alpha)$ is a modification factor of the Korringa relation, which depends on the exchange enhancement factor $\alpha$. If the spin fluctuations are enhanced around $q = 0$ as for dominant FM correlations, then $K(\alpha) < 1$. On the other hand, $K(\alpha) > 1$ indicates that finite-$q$ (typically AF) spin fluctuations are dominant. The estimated $K(\alpha)$ values for $x$ = 0.35 and 0.405 at 2 K (a temperature where the modified Korringa law is valid) are shown in Table I which indicates that the dominant ferromagnetic correlations are reduced with increasing $x$ or AF correlations are becoming dominant with increasing $x$ towards CeVGe$_3$.

\begin{figure}[h]
{\centering {\includegraphics[width=8.5cm]{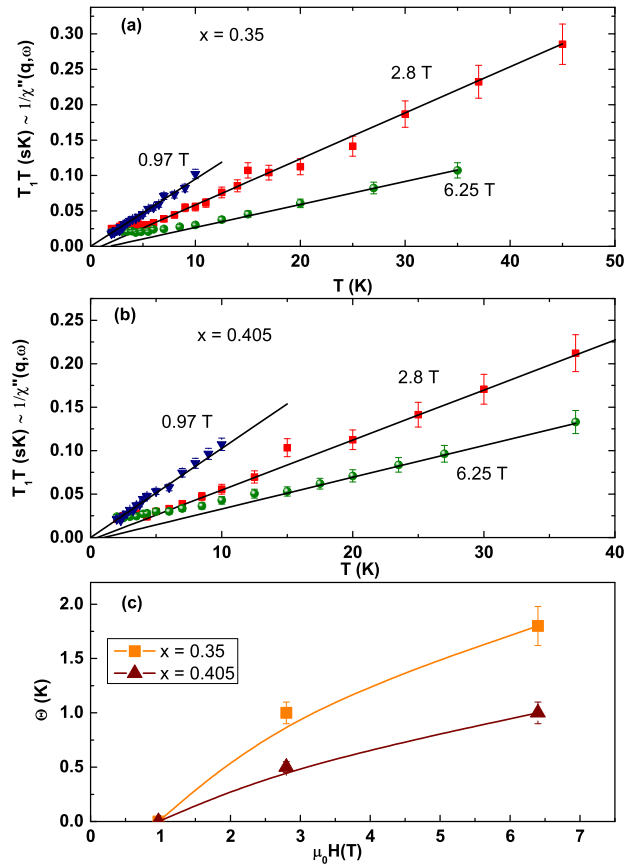}}\par} \caption{(a) and (b): Tempereature dependecne of $T_1T$ at different fields for $x$ = 0.35 and 0.405, (c): $\theta$ obtained from high temperatiure Curie-Weiss fits (straight lines) as a function of field for $x$ = 0.35 and 0.405. The curved lines are for guide to the eye.} \label{structure}
\end{figure}

Evidence for the presence of weak AF spin fluctuations on top of dominant FM spin fluctuations close to an FMQCP has been seen recently in Ru-doped CeFePO~\cite{Kitagawa12}. The appearance of AF spin fluctuations was atttributed to the Fermi-surface instability which might appear in case of a local QCP. The presence of considerable AF correlations for $x$ = 0.113 even far away from the possible QCP indicates that the QCP in this system is not driven solely by FM fluctuations. Furthermore, the Knight-shift anisotropy reduces considerably in CeTi$_{1-x}$V$_x$Ge$_3$ upon approaching QCP which indicates the isotropic nature of local fields at the V site. This behavior is very different from that of the layered FM CeRuPO, which is driven to a QCP by substitution of Ru by Fe, where the fluctuations become strongly anisotropic upon approaching the QCP~\cite{Kitagawa12,Kitagawa13}.

\begin{table}[h]
\caption{Summary of estimated parameters of CeTi$_{1-x}$V$_x$Ge$_3$.}
\begin{tabular}{l l l l l}
\hline \hline
CeTi$_{1-x}$V$_x$Ge$_3$ & x=0.113 & x=0.35 & x=0.405 &x=1\\
\hline
$T_0$ (K) & 22.9 & 21 & 11 & -\\
$1/T_1T$(1/sK) at 4K & 18.6 & 50 & 45 & 33\\
$A_{hf}^{iso}$ (T/$\mu_B$) & 0.9 & 0.6 & 0.5 & 1.1\\ 
$K(\alpha)$ & - & 0.28$\pm$0.02 & 0.46$\pm$0.05 & -\\
\hline\hline
\end{tabular}
\end{table}

On more general grounds, the issue of an FMQCP remains very challenging despite the considerable amount of work in this field~\cite{Brando}. It is generally believed that a QCP in clean systems is intrinsically unstable because of the dynamics of low-lying fermionic excitations. Therefore, alloy systems with varying degrees of disorder are an important subject. The system CeTi$_{1-x}$V$_x$Ge$_3$ is unique, in that FM for $x = 0$ gives way to AF for $x = 1$ and, at the same time, the single-ion anisotropy changes from uniaxial (Ising) to planar (XY). Close to the QCP, the system becomes isotropic. It should be mentioned that unusually slowly fluctuating glass-like electronic phases near FM quantum criticality due to the competing interactions have been proposed by theory~\cite{Nussinov09}. In our case, the competition between dominant FM and weak AF correlations might induce such phases which prevents the formation of a pure FMQCP. However, our NMR data do not indicate pronounced line broadening which would signify a glassy-like inhomogeneous state. Numerous experiments on clean three-dimensional FM metals have shown that, in contrast to the original Hertz-Millis-Moriya model of quantum criticality, the FMQCP is unstable, and FM metals undergo a first-order phase transition to the paramagnetic or to an incommensurate phase as predicted by theory~\cite{Chubukov04,Belitz94,Belitz97}. As a matter of fact, pure CeTiGe$_3$ under high hydrostatic pressure follows this scenario, with several intervening magnetic phases until the paramagnetic state is finally reached around 6 GPa. Under magnetic field a wing-like structure appears at high pressures, again as predicted by theory\cite{Belitz05} and observed in other clean systems\cite{Taufour16}. For an alloyed system as studied in the present work, with a considerable degree of disorder at the QCP at $x \approx 0.4$, one might expect a genuine second-order transition. Indeed, it was suggested on theoretical grounds that a first-order FM transition might be "tuned" continuously to a FMQCP by disorder~\cite{Sang14}. Further work is necessary to elucidate how these features compete or cooperate at the QCP. In this respect it is worth to mention that, our study indicates the presence of anisotropy in shift and which in principle should also induce anisotropy in the relaxation processes. The effect of anisotropy in general have not been included in our present study, especially in case of interpreting the nature of spin-fluctuations as it is not possible to estimate the relaxation in different directions without having single crystals. So we have mainly used the isotropic part of the Knight shift and the "average" $T_1$ by using the stretched exponential function which includes the distribution of $T_1$ due to the anisotropy and also disorder. The issues related to such anisotropy will be of interest for future studies by employing single crystals. 

\section{conclusion}
Systematic $^{51}$V NMR measurements have been performed on CeTi$_{1-x}$V$_x$Ge$_3$ with the end members showing ferromagnetic and antiferromagnetic order, for $x = 0$ and $x = 1$ respectively. NMR as a local probe provides informations about magnetic fluctuations across the phase diagram. The temperature dependence of $K$ and $1/T_1T$ in CeVGe$_3$ shows strong admixture of FM fluctuations to the dominant AF fluctuations. The temperature dependence of $1/T_1T$ for $x =0.113$ at 6.4 T can be well explained by self-consistent renormalization theory for itinerant ferromagnets. Around the critical concentration ($x = 0.35, 0.405$), quantum-critical spin fluctuations comprise weak but finite AF spin fluctuations admixed to FM spin fluctuations. The spin-fluctuation parameters $T_0$ and $K(\alpha)$ (the latter probing the relative strength of AF vs. FM spin fluctuations) have been estimated for $x = 0.35$ and 0.405. $K(\alpha)$ shows a considerable enhancement with $x$ indicating the growing importance of AF fluctuations towards the QCP. The critical samples lack the NMR finger print of a pure FMQCP, i.e., the $1/T_1T \sim T^{-4/3}$ NMR power law~\cite{Majumder16}. Hence, the general presence of both FM and AF fluctuations across the whole CeTi$_{1-x}$V$_x$Ge$_3$ is a constituting trait of this system.  Further work should elucidate if and how the changing single-ion anisotropies affect the quantum criticality in this system.  

\section{Acknowledgment}
We thank M. Brando, C. Geibel and B. Pilawa for fruitful discussions. Furthermore, we thank H. Rave, C. Klausnitzer and R. Hempel-Weber for technical support. 

\section{Appendix A: crystal structure}
The crystal structure of Ce(V,Ti)Ge$_3$ has been shown in Fig. 6. It crystallizes in the hexagonal perovskite (BaNiO$_3$-type) structure P63/$\textit{mmc}$ with a=b=6.306 (6.2744)$\AA$, c=5.6732 (5.882)$\AA$, $\alpha=\beta=90^\circ$, $\gamma=120^\circ$ for CeVGe$_3$ (CeTiGe$_3$). The crystal structure has only one crystallographic V (Ti), Ce, and Ge sites. Fig. 6(a) is the full view and Fig. 6(b) is the view along the c axis, showing a uniaxial symmetry of the V site transferred hyperfine interaction from the Ce sites.

\begin{figure}[h]
{\centering {\includegraphics[width=8.5cm]{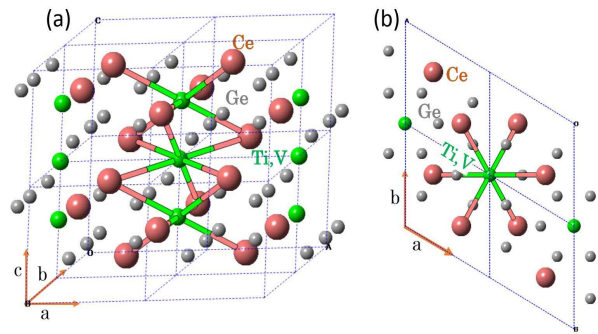}}\par} \caption{Crystal structure of Ce(Ti,V)Ge$_3$.} \label{structure}
\end{figure}

\section{Appendix B: Magnetization}
The dc-magnetization was measured in various magnetic fields and temperatures between 5 K and 100 K in temperature using commercially available SQUID magnetometer (Quantum Design MPMS). 

\begin{figure}[h]
{\centering {\includegraphics[width=8.5cm]{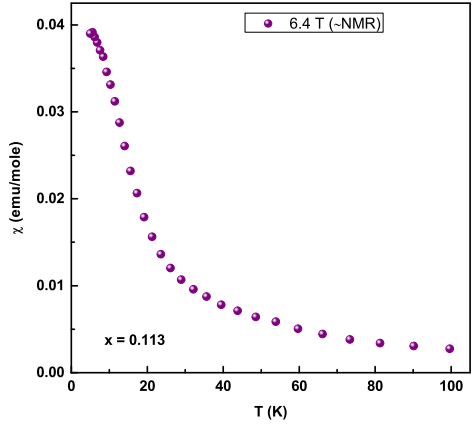}}\par} \caption{Temperature dependence of $M/H$ at $\mu_0$H = 6.4 T for the $x$=0.113 sample.} \label{structure}
\end{figure}

Fig. 7 shows the temperature dependence of ($\chi$ = $M/H$) for x=0.113 sample at 6.4 T close to which the $^{51}V$ NMR has been conducted.

Fig. 8 also shows the temperature dependence of $M/H$ and $\chi_{ac}$ (=$\delta M/\delta H$) at different magnetic fields in a log-log plot. We could not find any signature of long-range magnetic ordering. With increasing field the critical behavior of $\chi$ has been suppressed for the two samples close to the quantum-critical V concentration which is also consistent with the results of $^{51}V$ NMR described in the main text.

\begin{figure}[h]
{\centering {\includegraphics[width=8.5cm]{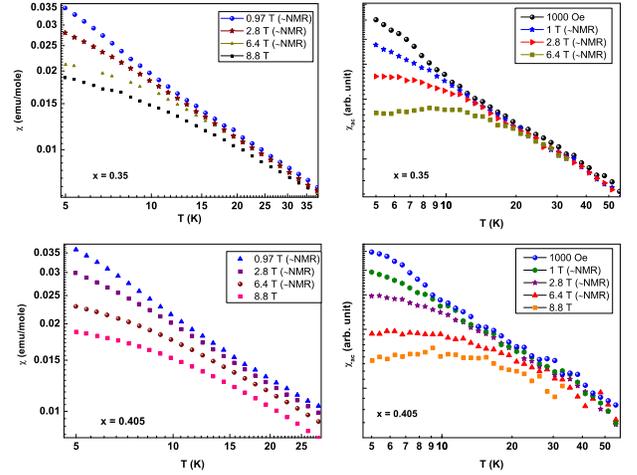}}\par} \caption{Temperature dependence of $M/H$ and $\chi_{ac}$ (=$\delta M/\delta H$) in different magnetic fields (mostly at which $^{51}V$ has been conducted) for the x=0.35 and 0.405 sample. The fields where $^{51}V$ NMR was performed are indicated.} \label{structure}
\end{figure}

\section{Appendix C: NMR}

The $^{51}V$ NMR spectra have been obtained using a commercially available Tecmag NMR spectrometer in the field sweep mode down to 1.8 K at various resonance frequencies (Fig. 9). NMR measurements have been performed on the $^{51}V$ nucleus with a spin (I=7/2) with 99.75\% natural abundance and a quadrupole moment of Q= -5.2$\times 10^{-30}m^2$.

\begin{figure}[h]
{\centering {\includegraphics[width=8.5cm]{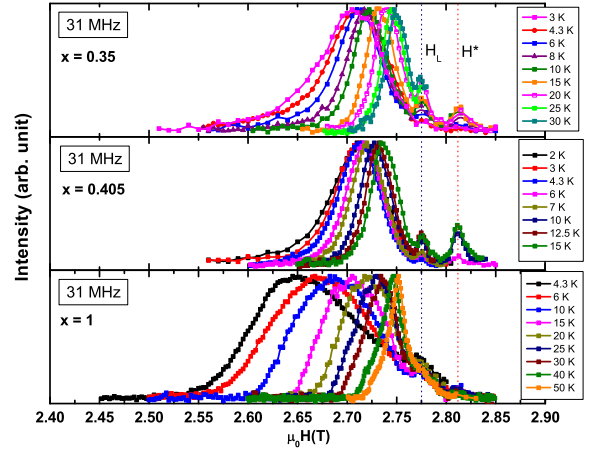}}\par} \caption{Field sweep $^{51}V$ spectra at 31 MHz. Vertical dotted lines at $H_L$ and $H^*$ (see text) are attributed to non-magnetic impurity phases.} \label{structure}
\end{figure}

The crystal structure of Ce(Ti,V)Ge$_3$ exhibits one V-lattice site. The $^{51}V$ NMR spectra shows only a single NMR line. There are no further lines (satellites) related to quadrupolar interaction. The spin-echo intensity was obtained by integrating over the spin echo in the time domain at a given magnetic field. The final spectrum is given by plotting the spin echo intensity as a function of the applied field. Due to the small quadrupolar moment and also small EFG at the V site the satellites are not visible, or they are superimposed with the broadened central line. The spectra for the $x$= 0.35 and 0.405 samples are quite isotropic in nature, whereas for $x$=0.113, the spectra are axially symmetric and for $x$=1 exhibit planar anisotropy. Interestingly the nature of anisotropy is opposite for both the samples ($x$=0.113 and 1), which we suggest is due to the opposite nature of anisotropy seen from magnetization measurements in single crystals~\cite{Inamdar14}. In addition, we have found two extra peaks in almost all spectra. One peak is at the $^{51}V$-Larmor field ($H_L$) which indicates the presence of unreacted V$_2$O$_5$. Another peak (at $H^*$) has a negative (but also temperature independent) shift which might originate from a binary $V^{5+}$-containing phase (probably V-Ge binary phase).

\begin{figure}[h]
{\centering {\includegraphics[width=8.5cm]{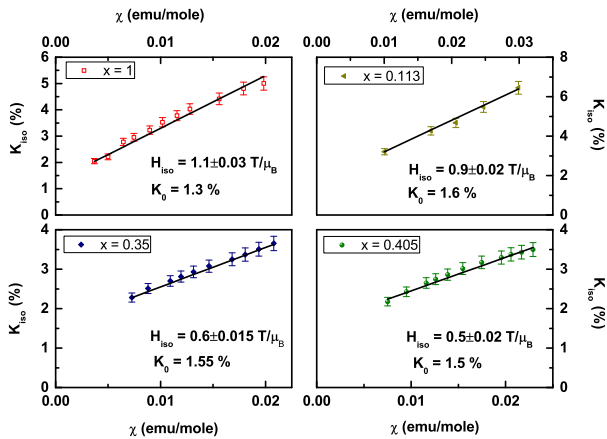}}\par} \caption{$K_{iso}$ versus $\chi$ for all the samples indicating a linear relation.} \label{structure}
\end{figure}

To estimate the hyperfine coupling constants $A_{hf}^{iso}$, $K_{iso}$(\%) is plotted versus $\chi$ (Fig. 10) and found to follow a linear relation as expected. From the slope we estimate $A_{hf}^{iso}$ which is plotted in Figure 4 in the main manuscript for all samples investigated.

\begin{figure}[h]
{\centering {\includegraphics[width=8.5cm]{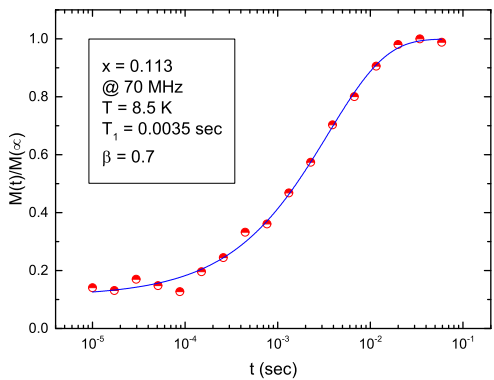}}\par} \caption{Magnetization recovery curve at 70 MHz and at 8.5 K for $x$=0.113 sample.} \label{structure}
\end{figure}

The spin-lattice relaxation rate was obtained by the standard saturation-recovery method. The exponent $\beta \approx$ 0.7-0.75 was kept constant for all temperatures for $x$=0.35, 0.405 and 0.113 but $\beta$=1 for the pure end member with $x$=1. A typical recovery curve has been shown in Fig. 11.

\begin{figure}[h]
{\centering {\includegraphics[width=8.5cm]{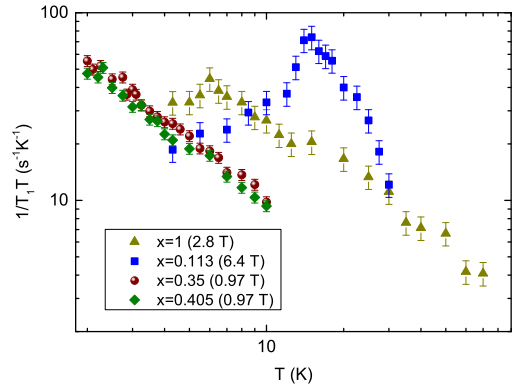}}\par} \caption{Temperature depedence of $1/T_1T$ at the respective lowest fields for each of the compounds. The figure indicates the doping evolution of electron spin-dynamics in CeTi$_{1-x}$V$_x$Ge$_3$.} \label{structure}
\end{figure}


\begin{thebibliography}{50}
\bibitem{Doniach} S. Doniach in "Valence instabilities and related narrow band phenomena" edited by R. D. Parks, page 1669 (Plenum New York 1977).

\bibitem{review} Hilbert v. L{\"o}hneysen, Achim Rosch, Matthias Vojta, and Peter Wolfle, Reviews of Modern Physics \textbf{79}, 1015 (2007).

\bibitem{Gegenwart08} P. Gegenwart  Q. Si and F. Steglich, Nat. Phys. \textbf{4}, 186 (2008).

\bibitem{Si10} Q. Si and F. Steglich, Science \textbf{329}, 1161 (2010).

\bibitem{Brando} M. Brando, D. Belitz, F. M. Grosche, and T. R. Kirkpatrick, Review of Modern Physics \textbf{88}, 025006 (2016).

\bibitem{Steppke13} A. Steppke, R. K\"{u}chler, S. Lausberg \textit{et al.}, Science \textbf{339}, 933 (2013).

\bibitem{Krellner07} C. Krellner, N. S. Kini, E. M. Br\"{u}ning \textit{et al.}, Phys. Rev. B. \textbf{76}, 104418 (2007).

\bibitem{Suellow99} S. S\"{u}llow, M. C. Aronson, B. D. Rainford, and P. Haen, Phys. Rev. Lett. \textbf{82}, 2963 (1999).

\bibitem{Sidorovat03} V. A. Sidorov E. D. Bauer, N. A. Frederick, J. R. Jeffries, S. Nakatsuji, N. O. Moreno, J. D. Thompson, M. B. Maple, and Z. Fisk, Phys. Rev. B \textbf{67}, 224419 (2003).

\bibitem{Sidorov05} V. A. Sidorov, E. D. Bauer, H. Lee, S. Nakatsuji, J. D. Thompson, and Z. Fisk, Phys. Rev. B \textbf{71}, 094422 (2005).

\bibitem{Lausberg12} S. Lausberg J. Spehling, A. Steppke, A. Jesche, H. Luetkens, A. Amato, C. Baines, C. Krellner, M. Brando, C. Geibel, H.-H. Klauss, and F. Steglich, Phys. Rev. Lett. \textbf{109}, 216402 (2012).

\bibitem{Westerkamp09} T. Westerkamp, M. Deppe, R. K\"{u}chler, M. Brando, C. Geibel, P. Gegenwart, A. P. Pikul, and F. Steglich, Phys. Rev. Lett. \textbf{102}, 206404 (2009).

\bibitem{Baenitz13} M. Baenitz, R. Sarkar, P. Khuntia \textit{et al}., Phys. Status Solidi C \textbf{3}, 540 (2013). 

\bibitem{Sarkar13}R. Sarkar, P. Khuntia, C. Krellner, C. Geibel, F. Steglich, M. Baenitz, Phys. Rev. B \textbf{85}, 140409 (2013).

\bibitem{Bruning08} E. M. Bruning, C. Krellner, M. Baenitz, A. Jesche, F. Steglich, and C. Geibel, Phys. Rev. Lett. \textbf{101}, 117206 (2008).

\bibitem{Kittler13} W. Kittler, V. Fritsch, F. Weber, G. Fischer, D. Lamago, G. Andre, and H. v. L{\"o}hneysen, Phys. Rev. B \textbf{88}, 165123 (2013).

\bibitem{Thesis} W. Kittler, Ph. D. thesis, Karlsruhe Institute of Technology, (2014). 

https://www.ksp.kit.edu/download/1000046409.

\bibitem{Kittler} W. Kittler, V. Fritsch, B.Pilawa, M. Baenitz, P. C. Canfield, and H. v. Löhneysen, to be published.

\bibitem{Udhara18} Udhara S. Kaluarachchi, Valentin Taufour, Sergey L. Budko, and Paul C. Canfield, Phys. Rev. B \textbf{97}, 045139 (2018).

\bibitem{Moriya85} T. Moriya, Spin Fluctuations in Itinerant Electron Magnetism (Springer-Verlag, New York, 1985).

\bibitem{Inamdar14} Manjusha Inamdar, A Thamizhavel and S K Dhar, J. Phys.: Condens. Matter {\bf 26}, 326003 (2014).

\bibitem{Moriya56} T. Moriya, Prog. Theor. Phys. {\bf 16}, 23 (1956).

\bibitem{Moriya63} T. Moriya: J. Phys. Soc. Jpn. {\bf18}, 516 (1963).

\bibitem{Majumder10} M. Majumder, K. Ghoshray, A. Ghoshray, B. Bandyopadhyay, and M. Ghosh, Phys. Rev. B {\bf 82}, 054422 (2010).

\bibitem{Moriya78} T. Moriya, Solid State Commun. \textbf{26}, 483 (7978).

\bibitem{Yasuoka} N. Inoue and H. Yasuoka, Solid State Comm. \textbf{30}, 341 (1979).

\bibitem{Kitagawa13} Shunsaku Kitagawa, Kenji Ishida, Tetsuro Nakamura, Masanori Matoba, Yoichi Kamihara, J. Phys. Soc. Jpn., \textbf{82}, 033704 (2013).

\bibitem{Kitagawa12} S. Kitagawa, K. Ishida, T. Nakamura, M. Matoba, and Y. Kamihara, Phys. Rev. Lett. \textbf{109}, 227004 (2012).

\bibitem{Nussinov09} Z. Nussinov, I. Vekhter, and A. V. Balatsky, Phys. Rev. B \textbf{79}, 165122 (2009).

\bibitem{Chubukov04} A. V. Chubukov, Catherine Pepin, and Jerome Rech, Phys. Rev. Lett. \textbf{92}, 147003 (2004).

\bibitem{Belitz94} D. Belitz, and T. R. Kirkpatrick, Rev. Mod. Phys. \textbf{66}, 261 (1994).

\bibitem{Belitz97} D. Belitz, and T. R. Kirkpatrick, Phys. Rev. B \textbf{56}, 6513 (1997).

\bibitem{Belitz05} D. Belitz, T. R. Kirkpatrick, and J\"{o}rg Rollb\"{u}hler, Phys. Rev. Lett. \textbf{94}, 247205 (2005).

\bibitem{Taufour16} V. Taufour, U. S. Kaluarachchi, R. Khasanov, \textit{et al}., Phys. Rev. Lett. \textbf{117}, 037207 (2016).

\bibitem{Sang14} Y. Sang, D. Belitz, and T. R. Kirkpatrick, Phys. Rev. Lett. \textbf{113}, 207201 (2014).

\bibitem{Majumder16} M. Majumder, M. Wagner-Reetz, R. Cardoso-Gil, P. Gille, F. Steglich, Y. Grin, and M. Baenitz, Phys. Rev. B \textbf{93}, 064410 (2016).

\end{thebibliography}
\end{document}